\documentclass{aa}
\usepackage{txfonts}
\usepackage{graphicx}
\usepackage{subcaption}
\usepackage{threeparttable} 
\usepackage[colorlinks=true,allcolors=blue]{hyperref}
\usepackage{gensymb}

\DeclareUnicodeCharacter{5415}{}  
\DeclareUnicodeCharacter{5EFA}{}  
\DeclareUnicodeCharacter{4F1F}{}  

\begin{document} 

\authorrunning{Mountrichas et al.}
\titlerunning{BH mass, galaxy mass, and halo environment}

\title{Black hole mass, host galaxy mass, and dark matter halos: Testing the environmental connection}

\author{G. Mountrichas\inst{1}, F. Shankar\inst{2}, F. J. Carrera\inst{1},  A. Georgakakis\inst{3} }
          
     \institute {Instituto de Fisica de Cantabria (CSIC-Universidad de Cantabria), Avenida de los Castros, 39005 Santander, Spain.
              \email{gmountrichas@gmail.com}
               \and
             Department of Physics and Astronomy, University of Southampton, Highfield, Southampton, SO17 1BJ, UK
              \and
              Institute for Astronomy \& Astrophysics, National Observatory of Athens, V. Paulou \& I. Metaxa, 11532, Greece
             }

\abstract{
We investigate the relation between supermassive black holes (SMBHs), their host galaxies, and large-scale dark-matter halos by combining broad-line X-ray AGN from the XMM--XXL and Stripe\,82X surveys with galaxies from VIPERS and SDSS/Stripe\,82. Building on the homogeneous host-galaxy catalogue developed in Paper~I, we test whether AGN with a given black-hole mass, $M_{\rm BH}$, occupy the same or different large-scale environments as non-AGN galaxies with statistically indistinguishable host properties. We characterise the empirical $M_{\rm BH}$--$M_{\star}$ distribution of the AGN sample. A shallow scaling between $M_{\rm BH}$ and stellar mass, $M_{\star}$, is present, but with large intrinsic scatter influenced by flux-limited selection and virial-mass uncertainties. The ratio $M_{\rm BH}/M_{\star}$ declines with increasing $M_{\star}$ over the sampled range. Overmassive and undermassive AGN subsets, defined relative to this empirical trend, exhibit distinct median host properties consistent with partially non-synchronous SMBH and $M_{\star}$ growth. We then select AGN in two $M_{\rm BH}$ intervals,
$8.0 \le \log(M_{\rm BH}/M_\odot) < 8.5$ and
$8.5 \le \log(M_{\rm BH}/M_\odot) < 9.0$,
and construct galaxy control samples matched in $M_{\star}$, SFR, and sSFR ($=\frac{SFR}{M_{\star}}$) using a multivariate nearest-neighbour procedure. Using AGN--galaxy cross-correlation functions, we infer characteristic halo masses for AGN and matched galaxies in each bin. AGN with
$8.0 \le \log(M_{\rm BH}/M_\odot) < 8.5$
occupy halos statistically indistinguishable from those of their controls, indicating no detectable environmental dependence at these masses once host properties are controlled. In the higher-mass bin,
$8.5 \le \log(M_{\rm BH}/M_\odot) < 9.0$,
we find a mild indication that AGN may reside in more massive halos than the matched non-AGN galaxies. The inferred difference is $\sim0.4$\,dex but remains formally consistent within the uncertainties. If confirmed with larger samples, this may indicate that halo-scale processes become increasingly relevant only at the highest $M_{\rm BH}$, while at lower masses AGN environments remain indistinguishable from those of inactive galaxies with similar host properties.
}

\keywords{}
   
\maketitle  

\section{Introduction}
\label{sec_intro}

Supermassive black holes (SMBHs) reside at the centres of most galaxies and their growth is intimately linked to the assembly of stellar mass, $M_\star$, and the evolution of their hosts. This connection is supported by a set of tight empirical relations between black hole mass, $M_{\rm BH}$, and bulge properties such as $M_\star$ or velocity dispersion \citep[e.g.][]{Magorrian1998, Marconi2003, Haring2004, Kormendy2013}. Understanding when, where, and under what conditions SMBHs grow therefore requires not only characterising their internal properties, but also the large-scale environments in which they reside. Clustering analyses offer a powerful diagnostic of this connection because they provide direct estimates of the typical dark matter halos (DMHs) hosting active galactic nuclei (AGN), thereby linking SMBH growth to cosmic structure formation.

Over the past two decades, extensive work has examined AGN environments across a range of redshifts and luminosities. X-ray AGN have played a central role in this effort owing to their relatively unbiased view of accretion and well-understood selection functions. Large-area surveys such as ROSAT, XMM–Newton, Chandra, and more recently eROSITA have enabled the measurement of AGN clustering over cosmologically representative volumes \citep[e.g.][]{Gilli2005, Coil2009, Krumpe2010b, Koutoulidis2013, Georgakakis2014, Mendez2016, Aird2019, Comparat2023}. A consistent picture has emerged in which moderate-luminosity X-ray AGN inhabit halos of approximately constant mass, $\log (M_{\rm halo}/M_\odot)\sim 12.5$--$13$, with surprisingly little evolution out to $z\sim3$ if not even higher \citep[][]{Allevato2011, Allevato2016}. At the same time, several studies have reported a dependence of clustering amplitude on X-ray luminosity \citep[$L_X$, e.g.][]{Koutoulidis2013, Krumpe2012, Powell2020}, suggesting that the most powerful AGN may reside in more massive, group-sized halos. These trends raise a fundamental question: are AGN Eddington-ratio ($\lambda_{\rm Edd}$) distributions largely independent of environment, implying a relatively simple connection between accretion rate and halo mass, or do $M_{\rm BH}$, accretion state, and host-galaxy properties introduce additional layers of environmental dependence?

Efforts to link SMBH mass to AGN clustering have so far yielded mixed results. Optical studies have sometimes reported a weak dependence of quasar clustering on $M_{\rm BH}$ \citep[e.g.][]{Shen2009}, whereas X-ray-selected AGN generally show little or no such trend \citep[e.g.][]{Krumpe2015, Krumpe2023}. In \citet{Mountrichas2026b} (hereafter Paper~I), we used the combined XMM–XXL and Stripe\,82X samples to explore the dependence of clustering on several SMBH properties, including $M_{\rm BH}$, $\lambda_{\rm Edd}$, and $L_X$, and found no significant variation of large-scale clustering (i.e. bias) with $M_{\rm BH}$ over the range $7.5 \lesssim \log(M_{\rm BH}/M_\odot) \lesssim 10$ probed by those data. This absence of a clear trend is consistent with theoretical and empirical work suggesting that any apparent connection between $M_{\rm BH}$ and halo mass is largely mediated by their mutual dependence on host-galaxy properties, particularly $M_\star$. A controlled comparison between AGN and non-AGN galaxies matched in host properties therefore provides a more direct way to test whether active SMBHs of a given mass trace environments that differ from those of inactive galaxies.

The observed correlation between $M_{\rm BH}$ and $M_{\star}$, despite its substantial intrinsic scatter and systematic uncertainties \citep[e.g.][]{Shankar2016}, provides a practical framework for addressing this issue. If AGN of different $M_{\rm BH}$ truly trace different stages of co-evolution with their hosts, then their environments may differ when compared with galaxies whose $M_\star$ correspond to the $M_{\rm BH}$ inferred from the scaling relation. Such comparisons have rarely been carried out, as they require: (i) homogeneous $M_{\rm BH}$ estimates for large numbers of AGN, with well-understood statistical and systematic uncertainties; (ii) matched galaxy samples spanning the same redshift and $M_\star$ ranges; and (iii) a clustering methodology that isolates host-driven differences from those induced by AGN selection.

In this work we bring together the XMM--XXL and Stripe\,82X AGN samples from Paper~I with high-quality host-galaxy measurements to construct a unified $M_{\rm BH}$--$M_{\star}$ relation, characterise its scatter, and examine the properties of AGN lying on, above, and below this relation. Using this relation as an empirical reference, we organise AGN within two black-hole mass intervals, $8.0\!-\!8.5$ and $8.5\!-\!9.0$, and construct control samples of non-AGN galaxies matched in redshift, $M_\star$, star-formation rate (SFR), and specific SFR (sSFR=SFR$/M_\star$).

Because $M_\star$ is tightly correlated with halo mass, matching in $M_\star$ and star-formation properties effectively controls for the primary dependence of galaxy clustering on host properties. This allows us to address a more specific question: whether AGN represent a random subset of galaxies at fixed host-galaxy parameters, or whether the presence of an accreting black hole of a given mass is associated with a systematic difference in large-scale environment.

This approach provides a complementary perspective to the $M_{\rm BH}$-based binning explored in Paper~I. Rather than comparing AGN of different $M_{\rm BH}$ to one another, we compare AGN to inactive galaxies with statistically indistinguishable host properties. Our aim is therefore not to measure a residual dependence of halo mass on $M_{\rm BH}$ at fixed $M_\star$, but to test whether the presence of active accretion itself is associated with a different halo environment once host-galaxy properties are controlled. Any difference in clustering would indicate that nuclear activity, at fixed $M_\star$ and SFR, is linked to halo-scale processes beyond those encoded in the global stellar-to-halo mass relation. Conversely, the absence of a difference would imply that AGN are environmentally indistinguishable from inactive galaxies once host properties are controlled.

The paper is organised as follows. Section~\ref{sec_data} describes the AGN and
galaxy datasets and outlines the clustering methodology. In
Section~\ref{sec:mbh_mstar_section} we present the $M_{\rm BH}$--$M_\star$
relation for our AGN samples, while Section~\ref{sec_clustering_results}
focuses on the clustering measurements and derived halo masses. A summary of our
main findings and conclusions is provided in Section~\ref{sec_summary}.

Throughout this paper, we adopt a flat $\Lambda$CDM cosmology with 
$\Omega_{\mathrm{M}}=0.315$, $\Omega_{\Lambda}=0.685$, $h=0.674$ (i.e.\ $H_0 = 67.4~\mathrm{km\,s^{-1}\,Mpc^{-1}}$) and $\sigma_8 = 0.811$,  
consistent with the \citet{Planck2020} cosmological parameters.

\section{Data}
\label{sec_data}

This work uses broad-line X-ray AGN from the XMM--XXL North and Stripe 82X surveys, together with galaxy samples from VIPERS and SDSS/Stripe~82.  
A companion analysis (Paper~I) investigates AGN clustering as a function of $M_{\rm BH}$, $\lambda_{\rm Edd}$, and $L_{\rm X}$ using the same datasets.  
Here we provide only a concise description of the properties relevant to the $M_{\rm BH}$--$M_\star$ analysis and refer the reader to Paper~I for full technical details, including survey characteristics, photometric coverage, and SED-fitting configuration.

\subsection{X-ray AGN}
\label{sec_agn}

The XMM--XXL North field \citep{Pierre2016} covers $\sim25\,{\rm deg}^2$ and contains $\sim2500$ X-ray sources with reliable spectroscopic redshifts \citep{Menzel2016, Liu2016}.  
$M_{\rm BH}$ for 1786 BLAGN1 were obtained via single-epoch virial estimators using H$\beta$, Mg\,\textsc{ii}, or C\,\textsc{iv}, following the calibrations of \citet{Shen2011}.  

The Stripe~82X survey \citep{LaMassa2013a, LaMassa2013b, LaMassa2015, LaMassa2024} spans $31\,{\rm deg}^2$ and incorporates both \textit{Chandra} and \textit{XMM-Newton} observations.  
In this work we use the DR3 catalog, which includes black hole mass measurements for 1297 Type~1 AGN derived from SDSS spectroscopy.  

For both fields we adopt the rest-frame $2$--$10\,\mathrm{keV}$ $L_X$ provided in the corresponding catalogs, which are not corrected for intrinsic absorption and are derived assuming a standard spectral model, as appropriate for broad-line AGN.

\subsection{Host-galaxy properties}
\label{sec_sed}

Host-galaxy $M_\star$ and SFR were derived through SED fitting using the same \texttt{CIGALE}-based methodology \citep{Boquien2019, Yang2020, Yang2022} employed in Paper~I (see also e.g., \citealt{Mountrichas2021c, Mountrichas2022a, Mountrichas2023a, Mountrichas2023c, Mountrichas2024a, Mountrichas2024b}).  
We require detections in the optical, near-infrared (NIR), and mid-infrared (MIR) bands \citep[e.g.,][]{Mountrichas2022c, Mountrichas2024c, Mountrichas2024d}; reduced $\chi^2_{\rm r} \leq 5$ \citep[e.g.,][]{Masoura2018, Masoura2021, Buat2021}; and consistency between best-fit and Bayesian estimates of $M_\star$ and SFR \citep[e.g.][]{Mountrichas2021b, Mountrichas2022b, Pouliasis2022, Koutoulidis2022}.  
After applying these quality filters, 584 AGN in XXL and 550 AGN in Stripe~82X remain with reliable host-galaxy parameters.

\subsection{Non-AGN galaxy samples}
\label{sec_galaxies}

For the XMM--XXL field, we use galaxies from the VIPERS PDR-2 release \citep{Garilli2014, Scodeggio2018}, which provides $\sim45{,}000$ reliable redshifts at $0.5<z<1.2$.  
After SED-quality cuts and removal of systems with significant AGN contribution (${\rm frac_{AGN}} > 0.2$), the remaining VIPERS sample contains $\sim 10{,}000$ galaxies \citep{Mountrichas2023d}.  

For Stripe~82, we use SDSS galaxies located within the X-ray footprint of the Stripe~82X survey.  
Applying the same SED-fitting and AGN-removal criteria yields $\sim 180{,}000$ galaxies with reliable host-galaxy parameters \citep[][]{Mountrichas2025a}.

For clustering, we restrict both galaxy samples to the sky regions overlapping the AGN footprint and to the redshift interval $0.5 \leq z \leq 1.2$.  
This results in 2428 galaxies (XXL) and 19\,141 galaxies (Stripe~82) contributing to the cross-correlation measurements. These samples serve as the tracer population for the AGN--galaxy cross-correlation analysis and as the parent pool from which the matched non-AGN control samples are constructed.

\subsection{Final AGN and galaxy samples}
\label{sec_final}

After enforcing both SED and spatial-overlap requirements, the final clustering samples contain 203 AGN and 2428 galaxies in the XMM--XXL field, and 245 AGN and 19\,141 galaxies in Stripe~82X (Table~\ref{table_numbers}).  
These subsets form the basis for the $M_{\rm BH}$--$M_\star$ relation, the classification of overmassive and undermassive AGN, and the environmental measurements presented in this paper.

\begin{table}[ht]
\centering
\caption{Final number of AGN and galaxies used in the clustering analysis after all photometric, quality, and redshift cuts, restricted to the overlapping AGN--galaxy regions.}
\label{table_numbers}
\begin{tabular}{lcc}
\hline\hline
Field & X-ray AGN & Galaxies \\
\hline
XMM--XXL      & 203 & 2\,428 \\
Stripe 82X    & 245 & 19\,141 \\
\hline
\end{tabular}
\end{table}

\subsection{Clustering analysis}
\label{sec_clustering}

We quantify the large-scale environments of AGN using the projected AGN--galaxy cross-correlation function, $w_p(r_p)$, following the methodology described in Paper~I \citep[see also][]{Mountrichas2009a, Mountrichas2012, Mountrichas2013, Mountrichas2016}. The cross-correlation approach reduces statistical uncertainties relative to the AGN autocorrelation and avoids the need to model the complex AGN selection function through random catalogues.

We stress that this methodology does not assume that AGN pairs are a subset of galaxy pairs, nor that AGN share the same halo occupation as inactive galaxies. The AGN bias is inferred from the AGN--galaxy cross-correlation relative to the galaxy autocorrelation in the linear two-halo regime, under the assumption of scale-independent bias on large scales. Any difference in halo occupation between AGN and matched galaxies would therefore manifest as a difference in the large-scale clustering amplitude.

The projected correlation function is obtained by integrating the redshift-space correlation function along the line of sight, adopting the same $\pi_{\rm max}$ values as in Paper~I. Uncertainties are estimated via Jackknife resampling, and covariance matrices are computed following \citet{Ross2008}.

We derive the AGN bias by fitting the measured $w_p(r_p)$ to the linear-theory dark matter prediction on scales $4<r_p<30\,h^{-1}{\rm Mpc}$, where the two-halo term dominates. Halo masses are then inferred using the \citet{Sheth2001} bias--mass relation with the \citet{Eisenstein1998} transfer function. As in Paper~I, we fit only the two-halo term of the correlation function, as the available AGN samples do not contain sufficient numbers of pairs to robustly constrain the one-halo term.

\subsection{Matched subsamples and parameter control}
\label{sec_matching}

To investigate whether AGN of a given $M_{\rm BH}$ inhabit different large--scale environments from non-AGN galaxies with comparable host properties, we compare AGN in specific $M_{\rm BH}$ intervals with carefully constructed control samples. A central challenge is that SMBH mass correlates strongly with several host-galaxy quantities, i.e., $M_\star$, SFR, and sSFR, all of which are known to influence galaxy clustering  
\citep[e.g.][]{Mountrichas2019, Allevato2019}.  
Our aim is to compare AGN in a given $M_{\rm BH}$ interval with non-active galaxies that share the same host-galaxy properties. Although the $M_{\rm BH}$--$M_\star$ relation provides a statistical mapping between $M_{\rm BH}$ and $M_\star$, the matching is performed directly on the measured $M_\star$, SFR, and sSFR of each AGN. In this way, the control galaxies reproduce the empirical host-property distributions of the AGN subsamples, rather than relying on the precise calibration of the global scaling relation.
By removing these dominant host-driven dependencies, the matching procedure allows us to isolate whether the presence of an actively accreting SMBH is associated with a distinct large-scale environment relative to inactive galaxies of similar stellar mass and star-formation properties. We emphasise that the clustering comparison is performed at fixed observed host-galaxy properties. Any uncertainty or intrinsic scatter in the $M_{\rm BH}$--$M_\star$ relation therefore does not directly propagate into the environmental measurement.

We employ the multivariate nearest-neighbour matching algorithm introduced in Paper~I.  
The method operates in a standardised multidimensional parameter space and performs one-to-one matching without replacement.  
A caliper of 1.2 (in normalised units) is applied to limit the maximum permitted distance between potential matches.  
This value provides an effective compromise between sample size and matching precision: smaller calipers reduce the number of usable AGN, while larger ones degrade the covariate balance.

Guided by our previous analyses of how environment depends on host-galaxy and SMBH properties, we match AGN to non-AGN galaxies in $\log M_\star$, $\log \mathrm{SFR}$, and $\log \mathrm{sSFR}$. Including the sSFR as an explicit matching variable improves control over the star-formation state of the galaxies in the presence of finite matching tolerances and intrinsic measurement scatter. This helps minimise residual environmental trends that correlate strongly with sSFR.

We experimented with simpler approaches, such as matching only on $M_\star$ or applying fixed threshold cuts, but these resulted in significantly smaller subsamples and poorer consistency across covariates.  
The multivariate matching adopted here therefore provides the most robust and statistically efficient framework for isolating environmental trends associated with SMBH mass.

\begin{figure}
\centering
  \includegraphics[width=0.9\columnwidth, height=6.5cm]{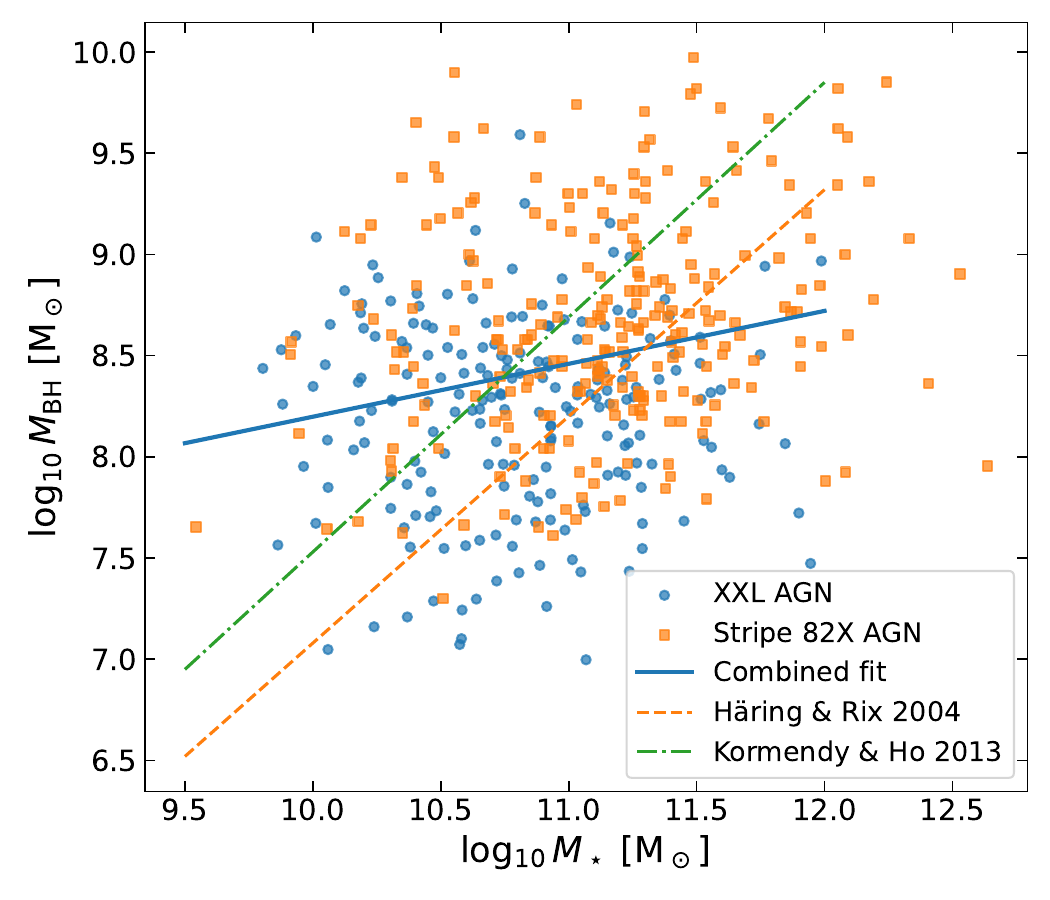} 
  \includegraphics[width=0.9\columnwidth, height=6.5cm]{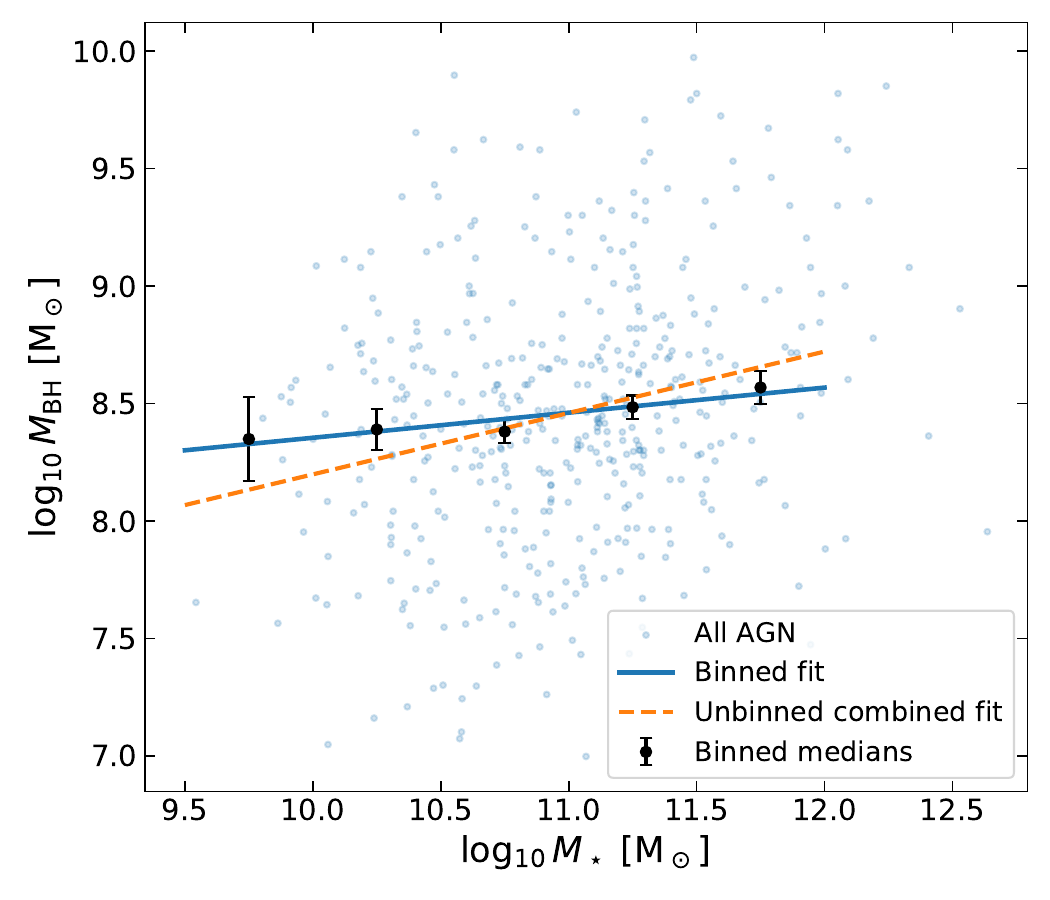}  
    \includegraphics[width=0.9\columnwidth, height=6.5cm]{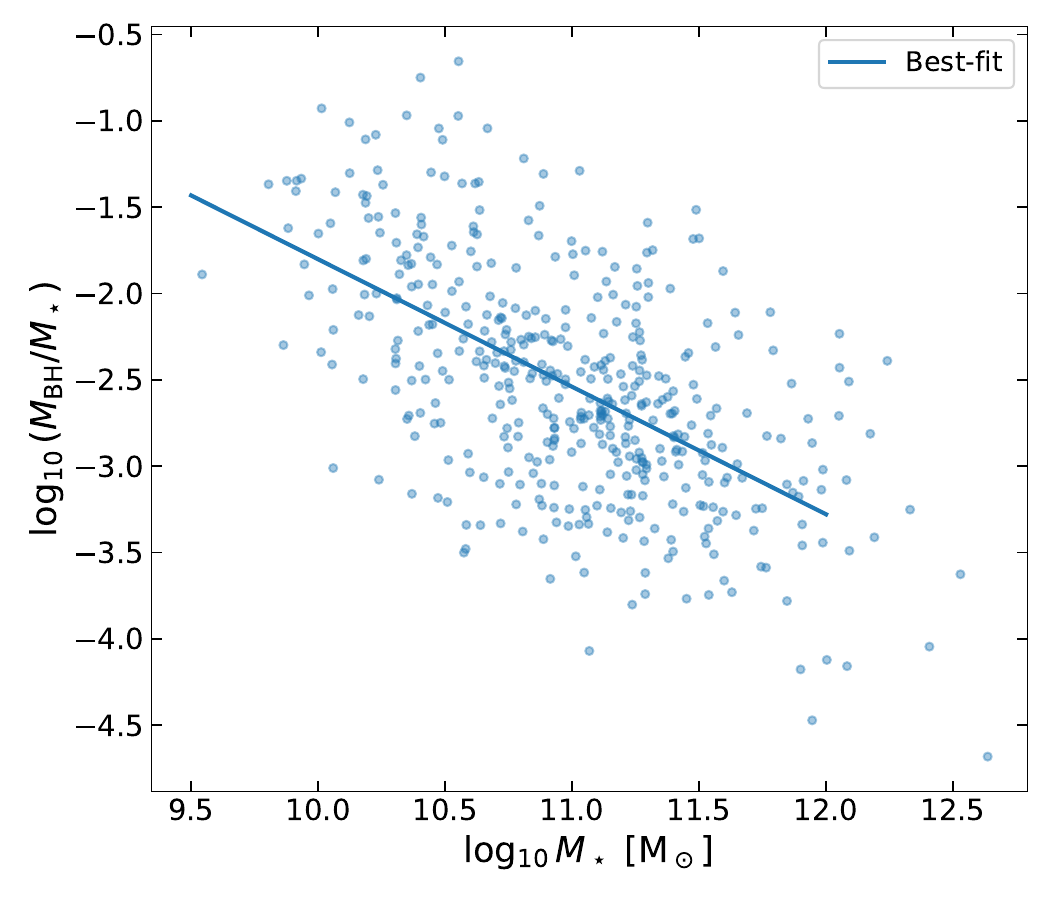}  
  \caption{Relation between $M_{\rm BH}$ and $M_\star$. Top: $\log M_{\rm BH}$–$\log M_\star$ relation for the XXL and Stripe\,82X AGN samples, together with the best-fitting linear trend for the combined dataset and the reference relations of Häring \& Rix (2004) and Kormendy \& Ho (2013). 
  Middle: Median $\log M_{\rm BH}$--$\log M_\star$ values in $M_\star$ bins (black points with bootstrap uncertainties), together with the corresponding binned fit and, for comparison, the unbinned combined fit from the top panel; the full AGN sample is shown in the background.
Bottom: $\log(M_{\rm BH}/M_\star)$ as a function of M$_\star$ for the combined AGN sample, with the best-fitting linear relation overplotted. }
  \label{fig:MBH_Mstar_panels}
\end{figure} 

\section{The $M_{\rm BH}$–$M_{\star}$ relation and definition of AGN and galaxy subsets}
\label{sec:mbh_mstar_section}

In this section, first we examine the $M_{\rm BH}$–$M_{\star}$ relation of our X-ray AGN samples. Then, we define the subsets used in the clustering analysis.

\subsection{The $M_{\rm BH}$--$M_{\star}$ relation}

We first examine the distribution of AGN in the $\log M_{\rm BH}$--$\log M_{\star}$ plane in order to characterise the empirical mapping between $M_{\rm BH}$ and $M_{\star}$ within our XMM--XXL and Stripe\,82X samples. Figure~\ref{fig:MBH_Mstar_panels} presents three complementary views:  
(i) the full AGN distribution in the $M_{\rm BH}$--$M_{\star}$ plane,  
(ii) median $\log M_{\rm BH}$ values in bins of $M_\star$, and  
(iii) the behaviour of the mass ratio $M_{\rm BH}/M_\star$ as a function of $M_\star$.

The top panel shows that the AGN populate a broad locus in the $M_{\rm BH}$--$M_{\star}$ plane, with substantial intrinsic scatter ($\gtrsim 0.4$\,dex) and no sharply defined ridge line. The distribution lies between the local dynamical relations of \citet{Haring2004} and \citet{Kormendy2013}, but exhibits a noticeably flatter trend across the stellar-mass range $10 \lesssim \log(M_\star/M_\odot) \lesssim 12$.

Fitting a linear model of the form
\begin{equation}
\log M_{\rm BH} = a + b\,(\log M_{\star} - 11)
\end{equation}
to the combined sample yields
\begin{equation}
a = 8.46 \pm 0.03,\qquad
b = 0.26 \pm 0.05.
\end{equation}
The fit is performed in $\log M_{\rm BH}$--$\log M_{\star}$ space using ordinary least squares, with uncertainties estimated via bootstrap resampling. We emphasise that this relation provides a statistical description of the present AGN sample rather than a calibration of the intrinsic SMBH--host scaling relation. Single-epoch virial masses carry typical uncertainties of $\sim0.3$--$0.5$\,dex, and the X-ray flux limit suppresses low-mass black holes at low $M_{\star}$, both of which can flatten the observed slope relative to local dynamical measurements.

When fitted separately, the two survey fields show modest but non-negligible differences in slope. 
For XXL we obtain a shallower relation, $b_{\rm XXL}=0.12\pm0.07$, whereas Stripe\,82X yields $b_{\rm S82X}=0.21\pm0.06$. 
This behaviour is consistent with the different dynamic ranges sampled by the two surveys: the XXL subset probes a narrower range in $M_{\rm BH}$ and is therefore more sensitive to the effects of scatter and sample truncation, while Stripe\,82X spans a broader range and recovers a somewhat steeper trend. 
This comparison illustrates that the inferred $M_{\rm BH}$--$M_\star$ scaling is sensitive to sample selection and mass coverage.

To better trace the central tendency, we compute median $\log M_{\rm BH}$ values in 0.5\,dex bins of $M_\star$ (middle panel). The corresponding binned fit yields
\begin{equation}
a_{\rm bin} = 8.46 \pm 0.02,\qquad
b_{\rm bin} = 0.17 \pm 0.03,
\end{equation}
slightly flatter than the unbinned result, as expected when medians reduce the influence of the upper envelope of overmassive systems. For comparison, the unbinned combined fit from the top panel is also shown in the middle panel. Together, these approaches indicate the presence of a broad but shallow positive correlation over the explored mass range.

The bottom panel shows the behaviour of the mass ratio:
\begin{equation}
\log(M_{\rm BH}/M_{\star}) = a_{\rm ratio} + b_{\rm ratio}\,(\log M_{\star}-11),
\end{equation}
with
\begin{equation}
a_{\rm ratio} = -2.54 \pm 0.03,\qquad
b_{\rm ratio} = -0.74 \pm 0.05.
\end{equation}
The negative slope largely reflects the shallow $M_{\rm BH}$--$M_\star$ trend discussed above. 
In the limit where $M_{\rm BH}$ varies only weakly with $M_\star$, a declining $M_{\rm BH}/M_\star$ ratio with increasing $M_\star$ follows naturally. Therefore, this panel does not provide independent information beyond the scaling shown in the upper panels, but offers an alternative representation of the same behaviour.

For the purposes of this work, the $M_{\rm BH}$--$M_\star$ relation serves primarily as a convenient empirical reference for organising the sample and defining relative deviations. The environmental analysis presented below is performed at fixed observed host-galaxy properties and therefore does not depend on the precise slope or normalisation of this fit.

\subsection{AGN classification and construction of galaxy control samples}
\label{sec:classification_controls}

We use the empirical $M_{\rm BH}$--$M_\star$ relation described above to organise the AGN sample for the clustering analysis. The primary goal of this paper is to compare AGN of a given  $M_{\rm BH}$ with non--AGN galaxies that share the same observed host-galaxy properties. For this reason, the analysis is defined first in terms of $M_{\rm BH}$ bins, rather than in terms of regions parallel to the $M_{\rm BH}$--$M_\star$ relation.

We focus on two $M_{\rm BH}$ intervals with sufficient statistics:
\[
8.0 \le \log(M_{\rm BH}/{\rm M_\odot}) < 8.5,
\qquad
8.5 \le \log(M_{\rm BH}/{\rm M_\odot}) < 9.0.
\]
These bins are chosen because they contain enough AGN for meaningful clustering measurements while still allowing us to examine whether AGN and inactive galaxies differ in environment at fixed black-hole mass.

Within each $M_{\rm BH}$ bin, we use the best-fitting global relation,
\begin{equation}
\log M_{\rm BH} = 8.46 + 0.26\,(\log M_\star - 11),
\end{equation}
as a reference to quantify whether individual AGN lie on, above, or below the mean $M_{\rm BH}$--$M_\star$ trend. We stress that this relation is not used to impose hard stellar-mass cuts on the galaxy sample. Instead, the clustering comparison is performed directly at fixed observed $M_\star$, SFR, and sSFR through the multivariate matching described below.

Given the substantial intrinsic scatter of the $M_{\rm BH}$--$M_\star$ relation ($\sim0.4$--$0.5$\,dex), we do not select galaxies based on a narrow predicted $M_\star$ range. Control samples are instead constructed via the multivariate nearest-neighbour matching scheme described in Sect.~\ref{sec_matching}. We perform 1-to-1 matching in standardised $\log M_\star$, $\log {\rm SFR}$, and $\log {\rm sSFR}$ space, ensuring that AGN and control galaxies share the same distributions in the host properties most strongly linked to halo occupation.

To characterise deviations from the mean relation, we define
\[
\Delta \log M_{\rm BH} =
\log M_{\rm BH} - \log M_{\rm BH,\,pred}(M_\star),
\]
and classify AGN as:
\begin{itemize}
\item On--relation: $|\Delta \log M_{\rm BH}| \le 0.3$\,dex,
\item Overmassive: $\Delta \log M_{\rm BH} > 0.3$\,dex,
\item Undermassive: $\Delta \log M_{\rm BH} < -0.3$\,dex.
\end{itemize}

The $\pm0.3$\,dex threshold is chosen to identify meaningful offsets from the mean trend while remaining comparable to typical single-epoch virial mass uncertainties \citep[e.g.][]{VestergaardPeterson2006, Shen2011}. This classification is intended primarily for qualitative interpretation rather than for clustering measurements.

For the clustering analysis, we retain only the on--relation AGN within each $M_{\rm BH}$ bin. The overmassive and undermassive subsets are too small for robust clustering: in the $8.0$--$8.5$ bin there are 31 undermassive AGN and no overmassive systems, while in the $8.5$--$9.0$ bin there are 45 overmassive AGN and no undermassive systems. Tables~\ref{tab:mbh_bin85_on_relation} and \ref{tab:mbh_bin90_on_relation} summarise the properties of the on--relation AGN and their matched controls, as well as of the overmassive and undermassive subsets.

The apparent asymmetry between bins reflects the shallow slope and intrinsic scatter of the empirical $M_{\rm BH}$--$M_\star$ relation combined with the selection of broad-line, flux-limited AGN, rather than a breakdown of the adopted scaling. Defining overmassive and undermassive systems globally across the full sample, without first separating in $M_{\rm BH}$, would mix objects with very different $M_{\rm BH}$ and would not address the central question of this work, namely whether AGN and inactive galaxies differ in environment at fixed $M_{\rm BH}$. Likewise, a percentile-based classification within $M_\star$ bins would produce bin-dependent definitions and would no longer isolate departures from a common global SMBH--host relation. We therefore adopt fixed $M_{\rm BH}$ bins for the clustering analysis and use the global relation as a common reference for defining relative offsets within those bins.

\subsection{Properties and interpretation of overmassive and undermassive AGN}
\label{sec:over_under_props}

Although the overmassive and undermassive subsets are too small for clustering, their median properties provide useful context for interpreting the scatter in the $M_{\rm BH}$--$M_\star$ plane. By construction, overmassive AGN lie above the mean $M_{\rm BH}$--$M_\star$ relation and therefore reside in hosts with lower $M_\star$ at fixed $M_{\rm BH}$. Importantly, these systems also exhibit elevated sSFR. In the high-$M_{\rm BH}$ bin, for example, the median $\log{\rm sSFR}$ of the overmassive subset is higher by $0.64$\,dex than that of the on--relation AGN ($0.26$ versus $-0.38$; Table~\ref{tab:mbh_bin90_on_relation}), indicating that their hosts are, on average, more actively star-forming.

\begin{figure}
\centering
\includegraphics[width=0.5\textwidth]{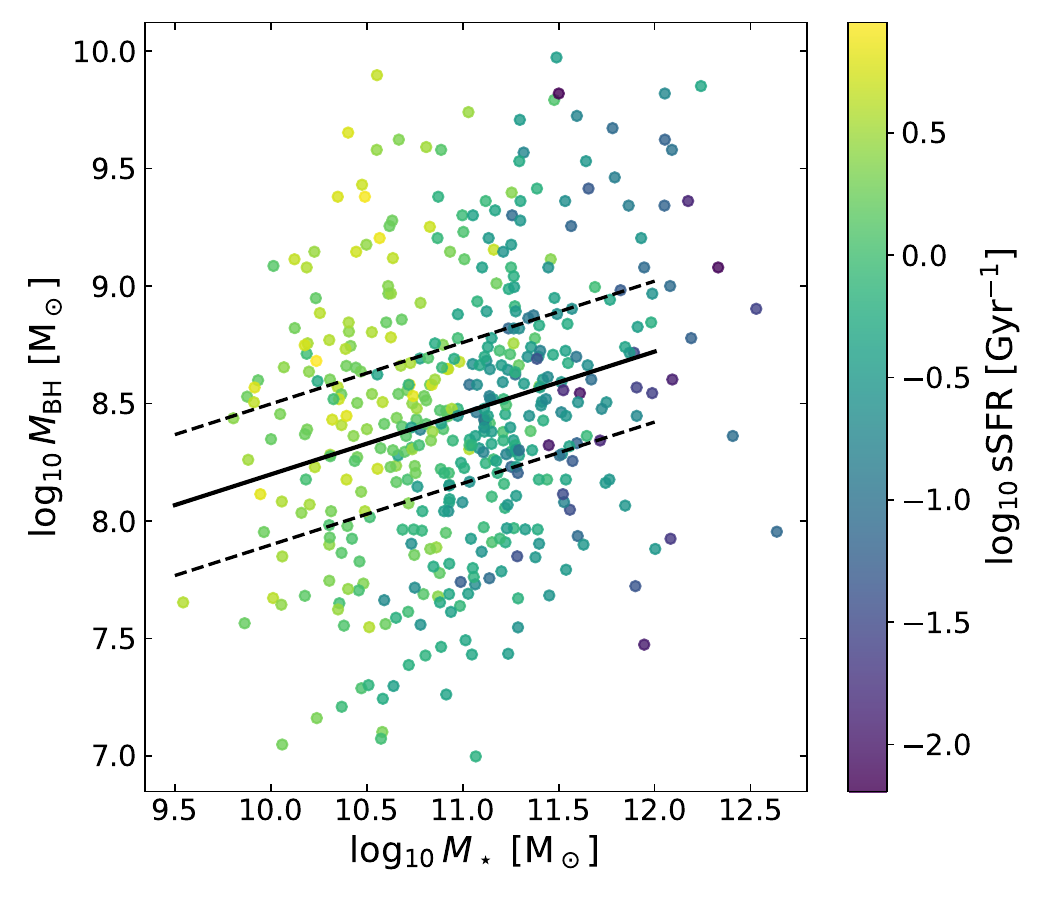}
\caption{Distribution of the AGN sample in the $\log M_{\rm BH}$--$\log M_\star$
plane colour-coded by $\log{\rm sSFR}$ (in ${\rm Gyr^{-1}}$). The solid black line
shows the best-fitting empirical $M_{\rm BH}$--$M_\star$ relation, while the dashed
lines indicate the $\pm0.3$\,dex thresholds used to define the on--relation,
overmassive, and undermassive subsets.}
\label{fig:mbh_mstar_ssfr}
\end{figure}

To visualise how star-formation state maps onto the $M_{\rm BH}$--$M_\star$ distribution, Fig.~\ref{fig:mbh_mstar_ssfr} shows the full AGN sample colour-coded by sSFR, together with the best-fitting relation and the $\pm0.3$\,dex boundaries used to define the three subsets. Higher-sSFR objects preferentially populate the low-$M_\star$ side of the relation at fixed $M_{\rm BH}$, consistent with the elevated sSFR measured for the overmassive subset.

However, we note that this behaviour may also reflect the well-known M$_\star$ dependence of galaxy star-formation activity. In particular, more massive galaxies are preferentially quiescent, while lower-mass systems tend to be actively star-forming. As a result, part of the observed sSFR gradient across the $M_{\rm BH}$--$M_\star$ plane may be driven by this underlying $M_\star$ sequence rather than by a direct dependence on SMBH mass. Indeed, at fixed $M_\star \sim 10^{11}\,M_\odot$, the AGN hosts in our sample broadly separate into star-forming and quiescent populations, with no obvious additional dependence on $M_{\rm BH}$.

This behaviour is in contrast with the trends reported by \citet{Terrazas2016, Terrazas2017}, who found that galaxies hosting overmassive black holes tend to have lower sSFR. In our sample, we observe the opposite behaviour, with overmassive AGN exhibiting higher sSFR. We attribute this discrepancy primarily to differences in sample selection. The Terrazas et al. studies focus on the general galaxy population, including quiescent systems, whereas our sample is restricted to actively accreting broad-line X-ray AGN. As a result, our selection preferentially includes systems with ongoing star formation and detectable accretion, while lower-sSFR, weakly accreting systems are underrepresented.

Consistent with this interpretation, even the on--relation AGN and their matched control galaxies in the higher-$M_{\rm BH}$ bin exhibit lower median sSFR than their counterparts in the lower-$M_{\rm BH}$ bin (Tables~\ref{tab:mbh_bin85_on_relation} and \ref{tab:mbh_bin90_on_relation}), indicating that the overall host population becomes less star-forming toward higher black-hole masses.

Undermassive AGN, conversely, occupy more massive and less star-forming hosts. In the lower-$M_{\rm BH}$ bin, the undermassive subset has a median $\log M_\star=11.28$, compared to $10.93$ for the on--relation AGN, and a median $\log{\rm sSFR}=-0.50$, lower by $0.50$\,dex than the on--relation population (Table~\ref{tab:mbh_bin85_on_relation}). What is not guaranteed by the definition, however, is how their instantaneous accretion rates compare to those of on--relation systems. In our sample, the undermassive AGN in the lower-$M_{\rm BH}$ bin have a median $\log \lambda_{\rm Edd}=-1.17$, higher by $0.19$\,dex than the on--relation AGN in the same bin. More generally, comparing the two extreme subsets across the two bins shows that the median sSFR differs by $0.76$\,dex and the median $\lambda_{\rm Edd}$ by $0.52$\,dex between the undermassive and overmassive populations (Tables~\ref{tab:mbh_bin85_on_relation} and \ref{tab:mbh_bin90_on_relation}). This suggests that the connection between star formation and accretion is non-trivial and not simply encoded by the position relative to the mean relation.

We quantified the link between star formation and accretion by testing for a correlation between sSFR and $\lambda_{\rm Edd}$. We find a weak but statistically significant correlation for the full AGN sample (Spearman $\rho\simeq0.22$), while the correlation is stronger within the on--relation and overmassive subsets ($\rho\simeq0.5$) and weaker for the undermassive population. This behaviour is consistent with partially coupled, but scatter-dominated, growth, in which episodes of enhanced accretion can coincide with elevated star formation but do not do so universally.

These contrasting behaviours motivate two schematic growth pathways, which we treat as qualitative interpretations rather than a deterministic evolutionary sequence:
\begin{itemize}
\item BH--leading episodes (overmassive):
  systems temporarily scattered above the mean relation, often associated with higher sSFR at fixed $M_{\rm BH}$.

\item Host--leading episodes (undermassive):
  systems temporarily scattered below the mean relation, for which comparatively high $\lambda_{\rm Edd}$ suggests that the SMBH may be in a phase of catching up.
\end{itemize}

This interpretation is consistent with our previous results. In \citet{Mountrichas2023b}, we showed that X-ray selected AGN tend to lie above the mean $M_{\rm BH}$--$M_\star$ relation on average, a natural consequence of flux-limited selection, which preferentially identifies systems with higher X-ray luminosities at fixed host-galaxy properties. We stress that the overmassive and undermassive classifications are statistical in nature: given the combined random and systematic uncertainties of single-epoch black-hole mass estimates ($\sim0.3$--$0.5$\,dex), the two subsets are expected to overlap substantially. As a result, individual sources may scatter across the adopted $\pm0.3$\,dex threshold without implying a true physical transition.

In \citet{Mountrichas2025a}, we further demonstrated in dwarf galaxies that $M_{\rm BH}/M_\star$ declines with increasing $M_\star$, consistent with early rapid black-hole growth followed by prolonged $M_\star$ assembly. The undermassive AGN identified here do not contradict this overall tendency. Rather, they are consistent with stochastic SMBH growth scenarios in which systems can temporarily scatter below the mean $M_{\rm BH}$--$M_\star$ relation and may subsequently experience phases of enhanced accretion. Related ideas have also been discussed by \citet{Terrazas2025} \citep[see also][]{Roberts2026}, who emphasised the diversity of black-hole growth histories and the importance of considering host-galaxy state when interpreting offsets from mean SMBH scaling relations. Given that the AGN and control samples are matched in $M_\star$, SFR, and sSFR, and that the observed accretion--star-formation coupling is scatter-dominated, this interpretation should be regarded as suggestive rather than conclusive.

A number of observational studies support the existence of host--leading episodes in which stellar mass assembly precedes significant SMBH growth. Analyses of SDSS AGN by \citet{Kauffmann2009} show that some massive, already-assembled galaxies host black holes that are undermassive relative to their stellar mass, consistent with delayed accretion episodes. Similar undermassive SMBHs have been identified in quenched or quenching systems \citep{Greene2020, Suh2020}, as well as in intermediate-mass galaxies where early star formation precedes late-time black-hole growth \citep{Li2021}. Studies of low-mass AGN also reveal undermassive black holes that may reflect early stellar assembly followed by a later phase of SMBH accretion \citep{Reines2015}. Together, these works support non-synchronous growth in which the host galaxy and SMBH can evolve on partially decoupled timescales.

The coexistence of overmassive and undermassive systems also agrees with predictions from cosmological hydrodynamic simulations \citep[e.g.][]{Habouzit2017, AnglesAlcazar2017, Dubois2021}, which show that SMBH and galaxy growth proceed in bursty, decoupled episodes that naturally produce large intrinsic scatter in the $M_{\rm BH}$--$M_\star$ plane. Semi-empirical and continuity-equation models \citep[e.g.][]{Shankar2013, Shankar2016, Shankar2020, Shankar2025, Aversa2016, Roberts2026} predict that duty-cycle effects, halo assembly history, and stochastic gas inflows generate large intrinsic scatter in the $M_{\rm BH}$--$M_\star$ plane, consistent with the coexistence of overmassive and undermassive systems in flux-limited AGN samples. Our findings therefore reinforce the view that the observed $M_{\rm BH}$--$M_\star$ distribution in AGN-selected samples reflects a combination of intrinsic diversity, stochastic growth, and selection effects.

\begin{table*}
\centering
\caption{Median properties of the on–relation AGN, their matched galaxy controls, and the overmassive and undermassive subsets in the
$8.0 \le \log(M_{\rm BH}/{\rm M_\odot}) < 8.5$ bin.}
\label{tab:mbh_bin85_on_relation}
\begin{tabular}{lcccc}
\hline\hline
Quantity & On–relation AGN & Control sample & Overmassive AGN & Undermassive AGN \\
\hline
Number of sources & 116 & 116  & --  &  31\\
$\log M_{\star}$ & 10.93 & 10.92 & -- & 11.28 \\
$\log{\rm SFR}$ & 1.83 & 1.83  & -- & 1.72\\
$\log{\rm sSFR}$ (Gyr$^{-1}$) & 0.00 & -0.02 & -- & -0.50 \\
$\log L_X$ & 43.90 & -- & -- & 43.88\\
$\log \lambda_{\rm Edd}$ & -1.36  & -- & -- & -1.17 \\
\hline
\end{tabular}
\end{table*}

\begin{table*}
\centering
\caption{Median properties of the on–relation AGN, their matched galaxy controls, and the overmassive and undermassive subsets in the
$8.5 \le \log(M_{\rm BH}/{\rm M_\odot}) < 9.0$ bin.}
\label{tab:mbh_bin90_on_relation}
\begin{tabular}{lcccc}
\hline\hline
Quantity & On–relation AGN & Control sample & Overmassive AGN & Undermassive AGN \\
\hline
Number of sources & 93 & 93 & 45 & -- \\
$\log M_{\star}$ & 11.25 & 11.25  & 10.47 & --\\
$\log{\rm SFR}$ & 1.92 & 1.90 & 1.78 & -- \\
$\log{\rm sSFR}$ (Gyr$^{-1}$) & -0.38 & -0.38 & 0.26 & --  \\
$\log L_X$ & 44.06 & -- & 44.00 & --\\
$\log \lambda_{\rm Edd}$ & -1.58  & --  & -1.69 & --\\
\hline
\end{tabular}
\end{table*}

\begin{figure}
\centering
  \includegraphics[width=0.9\columnwidth, height=6.5cm]{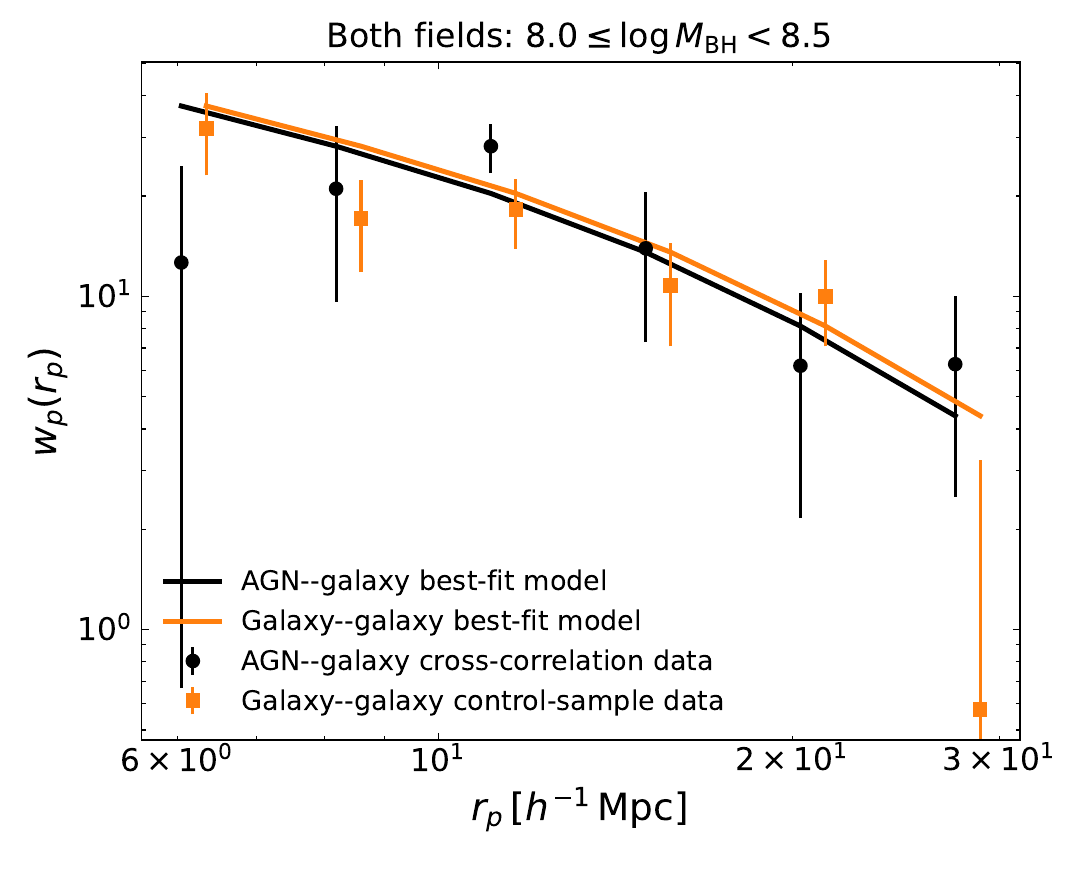} 
  \includegraphics[width=0.9\columnwidth, height=6.5cm]{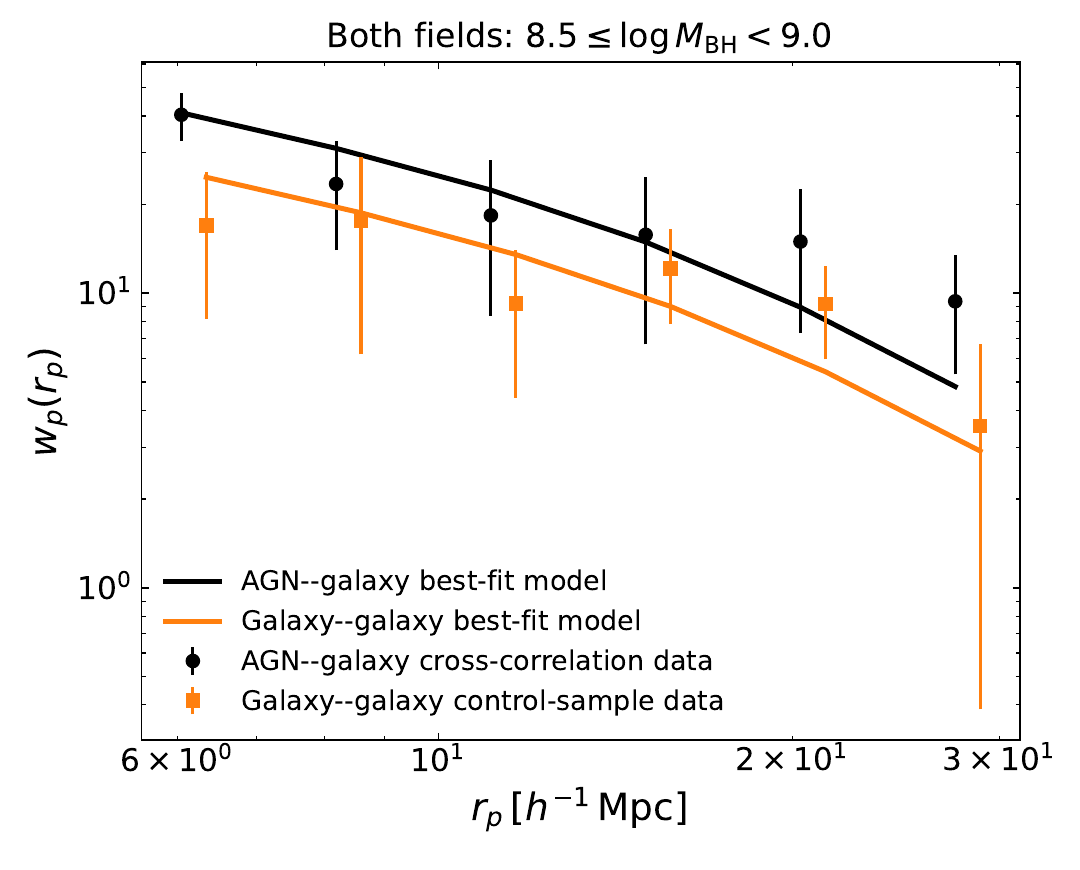}  
  \caption{Projected AGN--galaxy cross-correlation functions compared with the 
galaxy--galaxy cross-correlations of the matched control samples for the two 
black-hole mass intervals analysed in this work.}
  \label{fig_cluster_results}
\end{figure} 

\section{Clustering of AGN vs. non-AGN galaxies at fixed host-galaxy properties inferred from the $M_{\rm BH}$–$M_{\star}$ relation}
\label{sec_clustering_results}

In this section, we present the dark-matter halo (DMH) mass measurements for AGN and non–AGN galaxies that are matched in host-galaxy properties and discuss our findings.

\subsection{Clustering measurements}
First, we present the dark-matter halo (DMH) mass measurements for AGN and non--AGN galaxies that are matched in host-galaxy properties inferred from the $M_{\rm BH}$--$M_{\star}$ relation. 
The AGN halo masses are derived from the AGN--galaxy cross-correlation functions, using the galaxy bias inferred from galaxy--galaxy cross-correlations.

We estimate the environments of the matched galaxy control samples using galaxy--galaxy cross-correlation functions, adopting the same full galaxy population as tracer as in Paper~I. In this framework, the matched galaxies defined here are cross-correlated with the parent galaxy samples, ensuring that the tracer selection, random catalogues, and clustering methodology are identical to those used in Paper~I. For this reason, we do not repeat the methodological details or the corresponding measurements here. For reference, the DMH mass of the combined XXL+Stripe\,82X AGN sample was found to be 
$\log(M_{\rm DMH}/[h^{-1}M_{\odot}]) = 12.74 \pm 0.16$ (Paper~I).

Figure~\ref{fig_cluster_results} shows the AGN--galaxy cross-correlation functions alongside the galaxy--galaxy cross-correlation functions of the matched control samples, together with their best-fitting two-halo models.  
The top panel corresponds to AGN with $8.0 \le \log(M_{\rm BH}/{\rm M_\odot}) < 8.5$, and the bottom panel to AGN with $8.5 \le \log(M_{\rm BH}/{\rm M_\odot}) < 9.0$.  
All fits are performed over scales of $5 \le r_{\rm p} \le 30\,h^{-1}\,{\rm Mpc}$, since on smaller scales the number of AGN--galaxy pairs becomes too small for statistically meaningful constraints.

For the lower black-hole mass bin ($8.0$--$8.5$), the inferred halo masses of AGN and control galaxies are consistent within the uncertainties, with
$\log(M_{\rm halo}/{\rm M_\odot}) = 13.04\pm0.27$ for AGN and
$12.95\pm0.15$ for the sources in the control sample.
In the higher $M_{\rm BH}$ bin ($8.5$--$9.0$), AGN appear to reside in somewhat more massive halos, with
$\log(M_{\rm halo}/{\rm M_\odot}) = 13.53\pm0.37$
compared to
$13.16\pm0.27$ for the matched non-AGN galaxies.
Although the best-fitting halo masses differ by $\sim0.4$\,dex, the measurements remain consistent within the quoted uncertainties.
The projected correlation functions of AGN and control galaxies exhibit broadly similar amplitudes and shapes, with most data points overlapping within their error bars across the fitted scales.

We therefore regard the inferred halo-mass difference in this bin as marginal and sensitive to statistical fluctuations, rather than as a definitive detection.
Notably, both AGN and control galaxies in this regime inhabit group-scale dark-matter halos.
If real, the observed offset would suggest that AGN hosting the most massive black holes preferentially occupy the higher-mass end of the group-halo population relative to non-AGN galaxies matched in $M_\star$, SFR, and sSFR.
Given the current uncertainties, the relative halo-mass offset between AGN and control galaxies within the group regime should be regarded as tentative.

\subsection{Interpretation of the clustering results}
\label{sec:interpretation}

The halo-mass measurements presented in Fig.~\ref{fig_cluster_results} reveal two regimes. 
AGN with 
$8.0 \le \log(M_{\rm BH}/{\rm M_\odot}) < 8.5$
reside in halos with 
$\log(M_{\rm halo}/{\rm M_\odot})=13.04\pm0.27$, 
statistically indistinguishable from the 
$\log(M_{\rm halo}/{\rm M_\odot})=12.95\pm0.15$ 
measured for their matched non-AGN galaxies. 
Given that AGN and control galaxies are matched in $M_\star$, SFR and sSFR, quantities known to correlate with large-scale bias and halo occupation 
\citep[e.g.][]{Georgakakis2014, Aird2019, Mountrichas2019, Allevato2019}, this result indicates that, at moderate $M_{\rm BH}$, the presence of an actively accreting SMBH does not correspond to a distinct large-scale environment once host-galaxy properties are controlled.

In contrast, in the higher-mass bin 
$8.5 \le \log(M_{\rm BH}/{\rm M_\odot}) < 9.0$, 
we find a marginal indication of a higher characteristic halo mass for AGN compared to the matched non-AGN galaxies. 
The best-fitting values are 
$\log(M_{\rm halo}/{\rm M_\odot})=13.53\pm0.37$ 
for AGN and 
$13.16\pm0.27$ 
for the control sample. 
Although the difference is $\sim0.4$\,dex, the measurements remain formally consistent within the uncertainties, and the projected correlation functions show broadly similar amplitudes and shapes. 
We therefore regard this offset as suggestive rather than conclusive.

At face value, the emerging pattern is consistent with a scenario in which environmental effects become increasingly relevant only in the high--$M_{\rm BH}$ regime. 
Several studies have reported enhanced AGN fractions in group and cluster environments 
\citep[e.g.][]{Martini2007, Martini2009, Martini2013}, 
while more recent analyses and modelling 
\citep[e.g.][]{Georgakakis2023, Munoz2024} 
suggest that AGN triggering in dense environments cannot be fully explained by an $\lambda_{\rm Edd}$ distribution that is independent of halo mass. 
In this context, the mild halo-mass excess detected here may reflect the increasing importance of halo-driven processes at the high-mass end of the SMBH population.

To assess whether structural differences may contribute to this signal, we examined available DESI Legacy Survey morphological parameters. In particular, we considered the single-component Sérsic index (SERSIC), the semi-major-axis size proxy (SHAPE\_R), and the TYPE classification provided in the DESI catalogs \citep{Quilley2025, Romelli2025}. The Sérsic index traces the concentration of the surface-brightness profile, increasing from disk-dominated systems ($n \sim 1$) to bulge-dominated systems ($n \gtrsim 4$), while SHAPE\_R provides a model-independent estimate of the apparent size. The TYPE flag broadly classifies sources as PSF (compact or unresolved), disky, bulgy, or irregular systems.

DESI morphological information is available for 15 AGN and 113 matched galaxies in the $8.0 \le \log(M_{\rm BH}/M_\odot) < 8.5$ bin, and for 8 AGN and 85 matched galaxies in the $8.5 \le \log(M_{\rm BH}/M_\odot) < 9.0$ bin. The smaller number of AGN with reliable DESI measurements reflects the seeing-limited depth of the Legacy imaging and the compactness of some AGN hosts.

\begin{figure}
\centering
\includegraphics[width=\columnwidth]{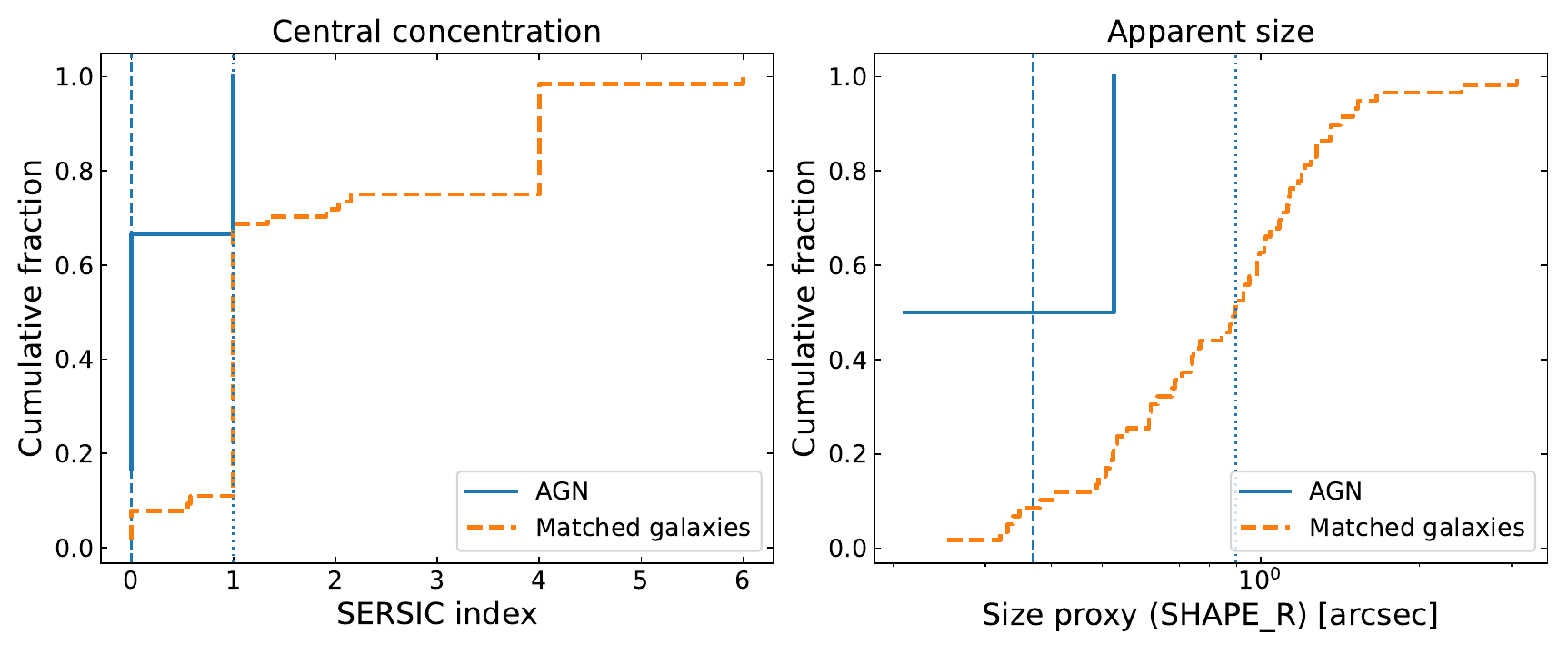}
\caption{Cumulative distributions of SERSIC (left) and SHAPE\_R (right) for AGN and matched control galaxies in the $8.5 \le \log(M_{\rm BH}/M_\odot) < 9.0$ bin. The vertical lines indicate the sample medians. Owing to the small number of AGN with reliable DESI morphology, especially for SHAPE\_R, this comparison is intended as a qualitative illustration only.}
\label{fig:desi_morph_highmbh}
\end{figure}

In the lower $M_{\rm BH}$ bin, no clear structural differences are apparent between AGN and matched galaxies, consistent with their indistinguishable halo masses. In the higher $M_{\rm BH}$ bin, we find a qualitative indication that AGN hosts tend to be more compact than their control counterparts. Figure~\ref{fig:desi_morph_highmbh} illustrates this comparison for the high-$M_{\rm BH}$ matched samples using the cumulative distributions of SERSIC and SHAPE\_R. The SHAPE\_R distributions suggest that AGN hosts occupy the smaller-size part of the distribution, whereas the Sérsic-index comparison is less conclusive, likely because unresolved nuclear emission can bias single-component profile fits. Given the limited statistics and the seeing-limited nature of the imaging, this comparison is necessarily qualitative. Nevertheless, it is consistent with the idea that, at fixed $M_\star$ and SFR, structural differences may correlate with both halo mass and SMBH mass in the high-mass regime.

Supermassive black holes above 
$\sim10^{8.5}\,{\rm M_\odot}$ 
are frequently associated with bulge-dominated or transitioning systems with high central densities 
\citep[e.g.][]{Kormendy2013, Greene2020, Ding2020}. 
Such galaxies preferentially inhabit group-scale halos where quenching and virial-shock heating become efficient at 
$M_{\rm halo}\sim10^{12.5}$--$10^{13}\,{\rm M_\odot}$ 
\citep[e.g.][]{Birnboim2003, Dekel2006}. 
If the tentative halo-mass excess observed here is real, it may therefore reflect the coupling between bulge assembly, halo potential depth, and SMBH growth that becomes prominent only at the massive end.

Finally, we emphasise that these results are fully consistent with Paper~I. 
Paper~I showed that among X-ray AGN, large-scale bias does not vary systematically with $M_{\rm BH}$ over the range probed. 
The present work addresses a different question: whether AGN occupy different halos than inactive galaxies with the same stellar and star-formation properties. 
The two results are therefore compatible. 
Within the AGN population, halo mass does not depend strongly on $M_{\rm BH}$; however, when high-mass AGN are compared to non-active galaxies at fixed host properties, a possible halo-related difference emerges only in the highest $M_{\rm BH}$ bin.

Hydrodynamic simulations 
\citep[e.g.][]{AnglesAlcazar2017, Habouzit2017, Dubois2021} 
and semi-empirical models 
\citep[e.g.][]{Shankar2013, Shankar2016, Shankar2020} 
predict that above 
$M_{\rm BH}\sim10^{8.3}$--$10^{8.5}\,M_\odot$, 
SMBH growth becomes increasingly sensitive to halo assembly and merger history, naturally producing enhanced scatter and mild environmental trends. 
Our measurements are consistent with this expectation, while remaining limited by current statistics. 
Future wide-area surveys with improved depth and morphology, such as \textit{Euclid}, will be required to determine whether the high-mass halo excess represents a genuine physical transition or a statistical fluctuation.

\section{Summary}
\label{sec_summary}

In this work we investigated the connection between SMBHs,
their host galaxies, and large-scale DMHs by combining
broad-line X-ray AGN from the XMM--XXL North and Stripe\,82X surveys with galaxy samples
from VIPERS and SDSS/Stripe\,82. These datasets were described extensively in
Paper~I, where we explored AGN clustering as a
function of SMBH properties. Here, we address a complementary question:
whether AGN with a given $M_{BH}$ occupy the same or different
large-scale environments as non-AGN galaxies with statistically indistinguishable
host-galaxy properties.

We first characterised the empirical $M_{\rm BH}$--$M_{\star}$ distribution of the
combined AGN sample. While a shallow scaling between $M_{\rm BH}$ and $M_{\star}$
is present in the data, the intrinsic scatter is large, and the observed
distribution is influenced by flux-limited AGN selection and virial-mass
uncertainties. The mass ratio $M_{\rm BH}/M_{\star}$ decreases with increasing
$M_{\star}$ within the sampled range, consistent with previous AGN studies and
with continuity-equation and semi-empirical models of SMBH growth
\citep[e.g.][]{Aversa2016, Shankar2016, Shankar2020}.
We stress that this relation serves primarily as an empirical characterisation
of the sample and as a bookkeeping device for defining subsamples, rather than
as a precise measurement of the intrinsic galaxy scaling relation.

Using this empirical relation, we classified AGN into on-relation,
overmassive, and undermassive subsets based on their deviations from the
best-fitting trend. Although the overmassive and undermassive subsets are too
small for clustering analysis, their median properties provide qualitative
insight into SMBH–host coevolution. Overmassive AGN reside in lower-$M_\star$
hosts at fixed $M_{\rm BH}$ and exhibit elevated sSFR, indicating ongoing
stellar mass assembly. Undermassive AGN occupy more massive, lower-sSFR hosts
and show comparatively high median $\lambda_{\rm Edd}$, suggestive of enhanced
accretion episodes. The correlation between sSFR and $\lambda_{\rm Edd}$ is weak
for the full sample but stronger within certain subsets, indicating partially
coupled, scatter-dominated growth rather than a one-to-one correspondence.
These trends are broadly consistent with stochastic, non-synchronous growth
scenarios predicted by simulations and semi-empirical models
\citep[e.g.][]{Aversa2016, Shankar2016, Shankar2020}.

For the clustering analysis, we selected AGN in two black-hole mass intervals,
\[
8.0 \le \log(M_{\rm BH}/{\rm M_\odot}) < 8.5, \qquad
8.5 \le \log(M_{\rm BH}/{\rm M_\odot}) < 9.0,
\]
and constructed galaxy control samples matched in $M_\star$, SFR, and sSFR
using the multivariate nearest-neighbour algorithm introduced in Paper~I.
This procedure ensures that AGN and non-AGN galaxies compared at fixed
black-hole mass share the same distributions in the host properties known to
correlate with halo occupation.

We measured AGN–galaxy cross-correlation functions for each black-hole mass bin
and compared them to the galaxy autocorrelation of the matched controls.
For moderate SMBH masses
($8.0 \le \log M_{\rm BH} < 8.5$), we find no measurable halo-mass difference
between AGN and inactive galaxies once host properties are controlled.
This indicates that, in this regime, the presence of an actively accreting SMBH
does not correspond to a distinct large-scale environment.

In the higher-mass bin ($8.5 \le \log M_{\rm BH} < 9.0$), we find a mild
indication of a higher characteristic halo mass for AGN relative to the matched
non-AGN galaxies. The inferred difference is $\sim0.4$\,dex but remains
formally consistent within the uncertainties. We therefore interpret this
signal as suggestive rather than conclusive. If real, it may indicate that
environmental or halo-scale processes become increasingly relevant only at
the highest black-hole masses, consistent with observational studies of AGN
in group and cluster environments \citep[e.g.][]{Martini2007, Martini2013}
and with recent modelling that suggests AGN triggering may depend on halo
mass beyond what is captured by an environment-independent $\lambda_{\rm Edd}$
distribution \citep[e.g.][]{Georgakakis2023, Munoz2024}.

Morphological information from DESI Legacy imaging suggests that, in the
high–$M_{\rm BH}$ bin, AGN hosts may be more compact than their matched
counterparts, while the evidence for enhanced central concentration is less
conclusive. The limited statistics and seeing-limited resolution imply that
this comparison remains qualitative.

These results are fully consistent with Paper~I, which found no detectable
dependence of AGN clustering on $M_{\rm BH}$ when comparing AGN to one another
at fixed host properties. The present analysis suggests that, within the AGN population, halo mass does not
vary systematically with $M_{\rm BH}$. However, when high-mass AGN are compared
to non-AGN galaxies at fixed host properties, a possible halo-related
difference emerges only in the highest mass regime.

Looking ahead, current and upcoming wide-area spectroscopic surveys, including
\textit{Euclid}, 4MOST, MOONS, WEAVE, the SDSS-V ``Black Hole Mapper'', and,
on longer timescales, the \textit{Roman} Space Telescope, will dramatically
increase the number of AGN and inactive galaxies with homogeneous measurements
of SMBH mass and host-galaxy structure. These datasets will reduce statistical
uncertainties, extend the dynamic range in $M_{\rm BH}$, and enable more
definitive tests of how SMBH growth, host-galaxy evolution, and halo
environment coevolve across cosmic time.

\begin{acknowledgements}
GM acknowledges funding from grant PID2021-122955OB-C41 funded by MCIN/AEI/10.13039/501100011033 and by “ERDF/EU”. This work was partially supported by the European Union's Horizon 2020 Research and Innovation program under the Maria Sklodowska-Curie grant agreement (No. 754510). This publication is part of the R\&D\&I project PID2024-155779OB-C31, funded by MICIU/AEI/10.13039/501100011033 and co-funded by FEDER, EU.  We also acknowledge partial support from the European Union’s Horizon 2020 research and innovation programme under the Marie Skłodowska-Curie grant agreement No 860744 (Bid4BESt; grant coordinator F. Shankar

\end{acknowledgements}

\bibliography{mybib}
\bibliographystyle{aa}

\end{document}